# A Framework for Data-Driven Physical Security and Insider Threat Detection

Vasileios Mavroeidis
Department of Informatics
University of Oslo, Norway
Email: vasileim@ifi.uio.no

Kamer Vishi
Department of Informatics
University of Oslo, Norway
Email: kamerv@ifi.uio.no

Audun Jøsang
Department of Informatics
University of Oslo, Norway
Email: josang@ifi.uio.no

*Abstract*—This paper presents *PS0*, an ontological framework and a methodology for improving physical security and insider threat detection. *PS0* can facilitate forensic data analysis and proactively mitigate insider threats by leveraging rule-based anomaly detection. In all too many cases, rule-based anomaly detection can detect employee deviations from organizational security policies. In addition, *PS0* can be considered a security provenance solution because of its ability to fully reconstruct attack patterns. Provenance graphs can be further analyzed to identify deceptive actions and overcome analytical mistakes that can result in bad decision-making, such as false attribution. Moreover, the information can be used to enrich the available intelligence (about intrusion attempts) that can form use cases to detect and remediate limitations in the system, such as loosely-coupled provenance graphs that in many cases indicate weaknesses in the physical security architecture. Ultimately, validation of the framework through use cases demonstrates and proves that PS0 can improve an organization's security posture in terms of physical security and insider threat detection.

## I. Introduction

Physical security is often overlooked when it comes to information security. In addition, most organizations are operating under the false assumption that all those who are granted internal access are necessarily trustworthy. An organization's *physical* security controls should be considered equally important and prioritized as its *technical* and *administrative (procedural)* security controls. Physical access to an organization's secure areas, systems and data can reduce the effect of the safety measures that are installed to ensure their confidentiality, integrity and availability, thus, making it easier for a malicious insider to perpetrate a crime. Common goals of a malicious insider include espionage, sabotage, theft, fraud, destruction of competitive advantage, corruption of critical data, and disclosure of sensitive information. Stolfo et al. [1] explain that there are three different vectors of insider attack: misuse of access, defense bypass, and access control failure. Verizon [2] reports that in average 25% of breaches in 2017 involved internal actors and 8% of the tactics used involved physical actions. Since physical attacks can be carried out with little or no technical knowledge we predict that the number of attempted physical attacks and breaches will increase in the future. An attacker only needs to identify one simple weakness to get a point of entry and potentially cause havoc. Therefore, it is important for organizations to identify and remediate vulnerabilities in both physical and digital spaces.

Insider threats could be categorized based on their intent as malicious or unintentional. A malicious insider deliberately attempts to access and potentially harm an organization. On the contrary, unintentional threat refers to situations in which damage occurs as a result of an insider who has no malicious intent.

Existing researches for insider threat detection focus on monitoring and analyzing user activity to detect potential system misuses and abuses in the digital space [3]–[15], with some of them having more theoretical underpinning, and only a few of them mentioning [16] or focusing on the physical space [17]. Specifically, [17] present an ontology developed for information security knowledge sharing by focusing on data center physical security compliance. To the extent of our knowledge our paper is the first one to present a holistic data-driven approach for physical security that can improve an organizations security posture against insider threats. Leaving the physical factor out of the data analytics for insider threat detection means that a massive attack vector is overlooked. For example, failures in the systems of critical societal infrastructures caused by physical attacks can rapidly lead to massive disruption in society and loss of lives. In addition, solely relying on traditional access control systems to thwart sophisticated attacks is inadequate, since they solely generate isolated alerts and offer limited context. In all too many instances, by the time security personnel manually determine if the alerts present a real threat, the attackers already have achieved their goals. Cybersecurity and physical security should complement each other for detecting and preventing malicious incidents that involve physical actions, accurately and in a timely manner. In addition, Capelli et al. [18] indicate that non-technical and technical indicators are equally important to insider threat detection and prevention. Recent work [19] introduced SOFIT, a knowledge base (ontology) of individual and organizational sociotechnical factors for insider threats that expands on ITIO [20] (ontology that focuses on describing technical/cyber events). The authors demonstrated through a use case that non-technical indicators can enhance proactive insider threat mitigation.

Physical security controls should not be treated as "mere" logging solutions, but being part of a larger strategy of ongoing security operations leveraging data-driven analytics that can provide the context and the forensic evidence needed for undeniable attribution, as well as enable proactive attack mitigation. Traditional physical security strategies for deploying physical access control systems do not take into consideration approaches for data collection, storage and processing, resulting in unintelligent solutions that trigger alerts only after an identified unauthorized activity. It is also a fact that an organization would need to utilize many resources to conduct forensic analysis after a breach for tracing the perpetrator (attribution) and reconstructing the attack, especially when deception techniques have been employed. Knowing what happened prior and after a successful compromise or attempted attack would allow an organization to improve its security strategy and policies, as well as proactively mitigate future attempts.

This paper presents a new ontological framework and a methodology that address the aforementioned limitations related to physical security and insider threat detection. Specifically, we present the *PS0 framework - an ontology for physical security* that can facilitate forensic analysis and proactively mitigate insider threats by using rule-based anomaly detection. In all too many cases, rule-based anomaly detection can identify employee deviations from organizational security policies. In addition, PS0 can be considered a provenance solution, because of its capability to reconstruct attack patterns based on the time-sequence between events. Provenance graphs can be further analyzed to identify deceiving actions and overcome bad decision-making, such as wrong attribution. Moreover, the information can be used to enrich the available intelligence (intrusion attempts) that can be used as use cases to detect and remediate limitations in the system, such as loosely-coupled provenance graphs that in many cases indicate weaknesses in the physical security architecture.

The rest of the paper is structured as follows. In Section II, we re-define physical security. Section III describes the steps towards effective physical security and explains how it is achieved by our framework (*PS0*) when converging information technology and physical security. In Section IV we introduce *PS0*, a new framework for effective physical security and insider threat detection. Section V presents the developed ontology that is used to conduct forensic data analysis and proactive insider threat detection. Validation and experimental results are provided in Section VI along with the forensic analysis and rule-based anomaly detection for a particular use case. Finally, Section VII concludes the paper.

## II. Defining Physical Security

*Definition: Physical security is the use of physical and procedural controls that are designed to effectively protect, deter, respond, and recover from intentional and unintentional events (e.g., fire, natural disaster, terrorism, vandalism, theft, espionage and sabotage) that could cause serious loss or damage against an organization's critical assets (e.g., people, data, facilities and systems).*

*Physical security controls* include the functions of prevention, deterrence, and detection through multiple layers of interdependent systems, such as entry controls, physical barriers, security guards, alarm and surveillance systems, motion detection, proximity cards, biometrics, etc.

*Procedural controls* refer to security practices that are used to mitigate identified risks by means of policies, procedures, and guidelines.

## III. Effective Physical Security

The reputational, financial and regulatory impact of having an organization's assets damaged, stolen, or disclosed can be extremely serious and can lead to bankruptcy. Insiders are difficult to detect since they belong to the organization and have access and knowledge pertaining to the location of critical assets and access controls. This amplifies when illegal activity involves collusive employees that can easily bypass procedural security controls or insiders that are unaware that are aiding a threat actor (malicious insider).

To mitigate risks and protect critical assets from insider threats, organizations should establish an effective insider threat program fully integrated into the risk management strategy taking into consideration the digital and the physical spaces, simultaneously. In addition, the application of physical security controls should be structured in layers that in many cases overlap. There is no single physical control that will fulfill all of an organization's security needs.

Effective physical security, like cybersecurity, requires actions for identifying, analyzing, evaluating and mitigating risk (risk management program). Techniques for risk identification include brainstorming, interviews, checklists, statistics, historical data and use cases, and modeling techniques, such as attack trees and threat modeling. Our system supports the risk identification process and overall the risk management program by following a systematic iterative information enriching approach that allows ingestion of information derived from forensic analysis after unsuccessful or succeeded attacks. The risk indicators are analyzed collectively in order to uncover hidden relationships and thwart deception. The identified patterns after risk analysis and evaluation should be used to improve the overall insider threat program (physical and logical-technical security).

Convergence of IT and physical security with analytics should holistically work towards common objectives following a risk management program and a clearly defined security policy. For example, efforts to secure access to databases, e-mail, and organizational networks are merging with access control and surveillance systems and all together are part of an analytics approach to deliver actionable information and stimulate data-driven decisions.

## IV. Framework for Physical Security and Insider Threat Detection (PS0)[1]

*PS0* (Figure 1) is an ontological framework with provenance capability for improving physical security and insider threat detection. This is achieved by detecting and analyzing complex attacks that involve physical actions that in many cases are difficult to detect by traditional means. A key-point of the technology is that it considers damaging incidents or events in their entirety, thus, providing the appropriate visibility and context needed for an organization to improve its security posture and reduce risk against insider threats. This is attained by adopting a systematic iterative information enriching approach that allows, firstly to improve the security strategy by identifying and mitigating vulnerabilities and secondly to improve the rule-based system that tracks policy compliance and suspicious activities. Successful implementation of PS0 relies on early-adopted risk management practices for deploying the appropriate physical, procedural and monitoring controls providing the required level of security and visibility to an organization's critical assets and environment. Specifically, *PS0* addresses the following questions:

- How and when a compromise or attempt occurred
- What are the goals of the attack and potentially its impact
- Is there any activity violating any security policies that could potentially be an indication of the early stages of an attack or an unintentional insider
- Who is responsible for the compromise, attempt, or policy violation

*PS0* (Figure 1) adopts a layered deployment approach, where related functionality is grouped into a common layer that provides simple interfaces towards other layers and components, thereby abstracting the internal design and structure. The rest of the section introduces the components of *PS0*.

- **Data Sources**: includes any output from deployed security components, systems and available devices that after processing (ontology compliant triples) is used from our system to facilitate forensics, policy compliance, and detecting abnormal activity. We distinct the data sources in two categories.
  - **Physical Space:** includes logs deriving from access control systems, such as RFID readers, numpads, motion sensors, cameras, biometrics and more.
  - **Systems:** includes logs deriving from any solution monitoring or protecting the digital space. This can be aggregated and analyzed logs from endpoints (e.g. sysmon), active directory, networks (e.g. NetFlow), access points (network access), printers, as well as other IP-based devices (BYOD).

  Portable IP-based devices can play a major role in insider threat detection and forensics (attack pattern reconstruction, attribution and deception) when logging their interaction with an organization's network access points and wireless beacon monitoring devices. This is achieved by having a clear mapping between the media access control address of a device, the user the device belongs to, and the media access control addresses of the access points in addition to the timestamp of the authentication and the duration of the connection. Every new device seen for the first time in the network is mapped to a user profile based on the credentials used to access that network.

- **Log Collection and Aggregation:** includes the consistent collection of data from their respective data sources and their aggregation to an intermediary server for storage and processing. After the collection, aggregation, and processing of data (parsing engine) it is forwarded to the core of the framework; the ontology. The intermediary server should be able to maintain the confidentiality, integrity and availability of the data. Unauthorized data deletion or alteration would result in missing or wrong provenance.

- **Parsing Engine**: this component is responsible for manipulating each output type to bring it to the appropriate and consistent format (chunks of triples) based on the ontology schema and the restrictions specified. The parsing engine will forward the created triples to the ontology.

- **Ontology:** is the main component of *PS0* and in its current state provides structured and unstructured information about the infrastructure, physical security, user profiles and user activity of an organization. The knowledge base is used to facilitate forensics that can lead in provenance graphs and attack attribution, to provide a clear topology of an organizations infrastructure, critical assets and stakeholders, as well as to ensure policy compliance and detect abnormalities indicating threat likelihood, through well-defined rules. Details about the development of the ontology per se can be found in Section *V* and on Github[1].

- **Forensic Analysis:** includes the process of issuing SPARQL queries to investigate a breach or to examine abnormal activities and policy violations (aided by the rule-based anomaly detection component). Early detection and swift investigation are critical to avert attackers and responding to threats. Inadequate or missing forensic data would result in loosely coupled provenance graphs (missing links) and consequently reduced visibility (which indicates insufficient security measures). In response to that, *PS0* forensics can be used to identify weaknesses and strengthen an organization's security strategy. Available analyzed information (intelligence) from incidents can serve as use cases to validate an updated security strategy and potentially being shared to help other organizations improve their security posture. Overall a successful forensic analysis should provide complete attack pattern, attack context and rich intelligence, infrastructure-wide visibility, and insights gained from front-line experience investigating complex attacks.

- **Rule-Based Anomaly Detection:** traditional access control systems are vulnerable to social engineering and can be easily bypassed by insiders. Rule-based anomaly detection focuses on detecting security policy violations and suspicious activity that in many cases can easily go unnoticed until an incident occurs. Detected suspicious activity or policy violations are classified in the ontology ("Incident" class) based on their threat level ("red", "amber" and "green") and can be further analyzed to uncover malicious intent, an attack in progress, or negligent employees that can be potent attack vectors. The classification is achieved through different rules or restrictions categorizing incidents in their respective threat level class. For the above requirements we created OWL restrictions and we additionally used the semantic web rule language (SWRL). OWL restrictions and rule-based reasoning is an important feature of ontologies that can be used to infer new information, as well as for classification. A sample of rules that would flag threat alerts is described right below. The rules should always be aligned to an organization's security policy.

---
[1] *PS0* Github repository: https://github.com/securitylab-ch/ps0

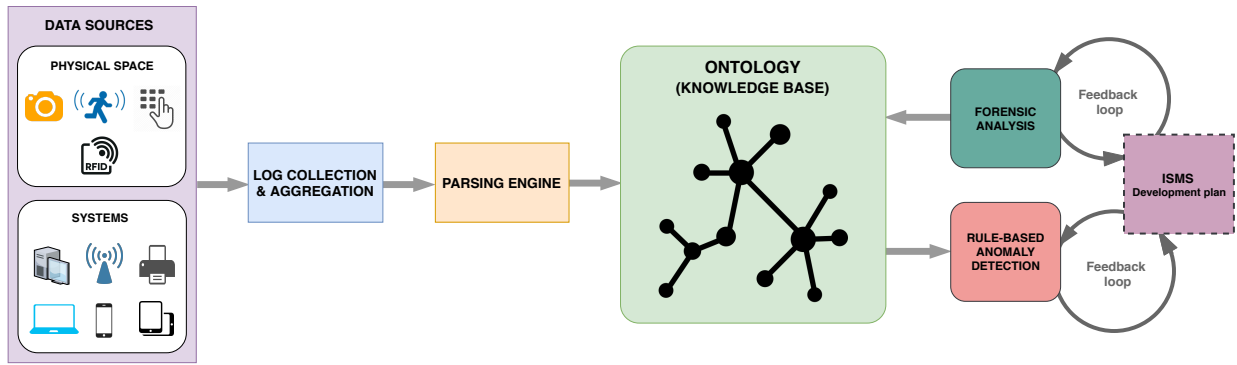

Fig. 1: High-level design of *PS0* framework.

- An individual was logged accessing one of the organization's secure environments (e.g., office, workspace, data-center) or one of the organization's systems (e.g., endpoint device, local area network) but was never logged entering the building or getting access to a particular access controlled area.
- An individual was logged entering a building but was never logged leaving the building in a specific time period (the opposite also applies).
- An individual was logged accessing the building more than one times but we do not have information of this individual leaving the building.
- Unauthorized user tried to gain access to a system or privileged credentials were used on regular workstations or servers which is considered policy violation.
- Events or incidents regarding access control to secure environments can be classified as follows. Three consecutive unsuccessful attempts (wrong PIN or passphrase) to an environment from the same user are classified as of high threat (red). The same applies to unsuccessful attempts that are not followed by a successful attempt in a specific time period. Two missing attempts followed by a successful attempt in a very short period is categorized as of low threat (green). Other areas use RFID technology for access control. Failed RFID attempts can be considered of low threat (green).

- **ISMS Development Plan:** Information Security Management System (ISMS) is a set of policies, procedures, technical and physical controls to protect the confidentiality, availability and integrity of an organization's sensitive data. An ISMS enables organizations to be significantly more resilient to both external and internal cyber and physical attacks and is responsible for coordinating cybersecurity with physical security. *PS0* should be part, aligned, and support the improvement of an organization's ISMS.

### V. PHYSICAL SECURITY ONTOLOGY

Part of our work was the development of an ontology capable of consuming and representing information coming from disparate data sources for conducting physical security forensic analysis, insider threat detection, and policy compliance. The ontology can be modified and adjusted taking into consideration each organization's unique infrastructure. The rest of the section introduces the most important classes of the ontology.

*Person* class includes detailed information of every entity associated with an organization. Information include personally identifiable information, associated roles, access rights and many more.

*GroupTeam and Department* classes include information about the organizational structure of an organization. A group is a collection of individuals who coordinate their individual efforts. A team is a group of people that come together to achieve a common goal. In an incident with more than one threat actors it is probabilistically possible that the accomplice would be member of the same group or team.

*RFID card* class includes information about all the available RFID cards distributed among stakeholders. In our case RFID cards grant access to different levels and workspaces in a building.

*MAC Address* class stores information about all the IP-based devices that have been observed connecting to a network. In addition, each MAC address is mapped to a specific

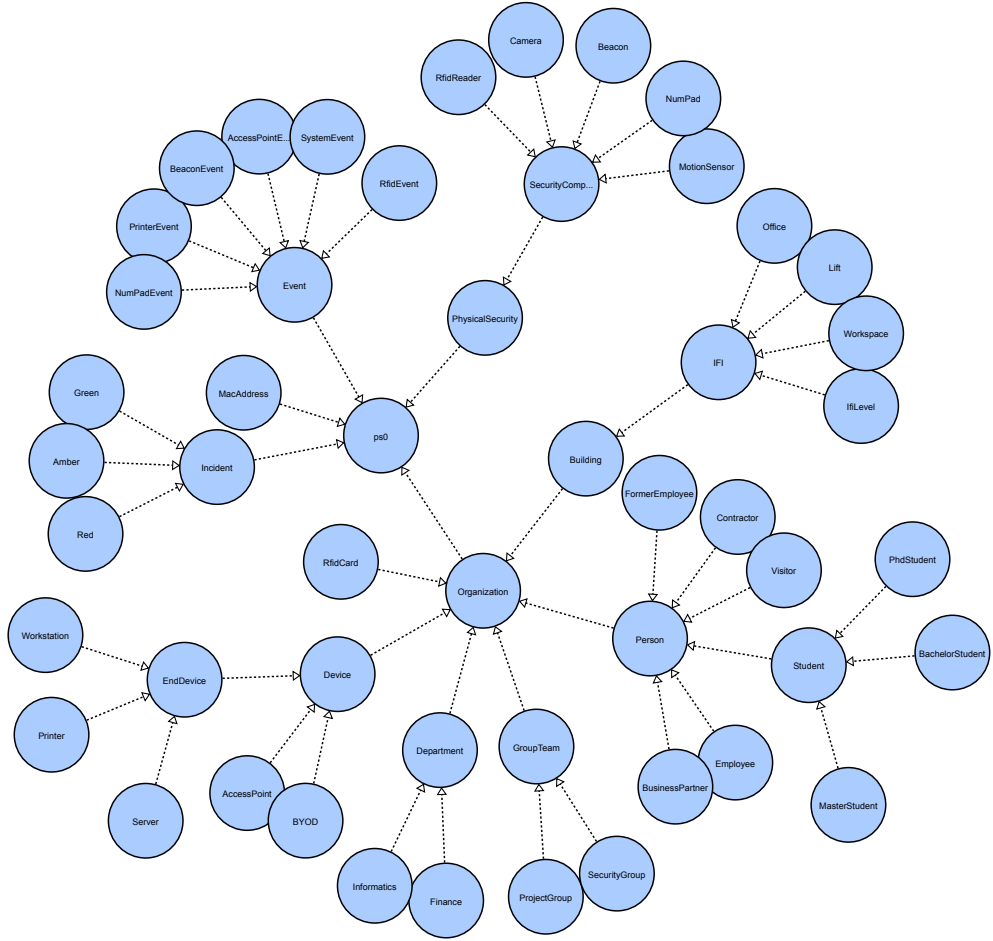

Fig. 2: Main classes of the ontology.

person. Each person has its own credentials to connect to the organizational network.

*Building* class maps a building (infrastructure) and enumerates the exact location and details of every office, workspace, security component and device.

*Event* class categorizes all the collected and aggregated logs coming from the installed security components and systems.

*Security Component* class includes all the security solutions installed and deployed (physical and logical controls) as well as information about them. This class is tightly connected with the class *Building* mapping the access control systems.

*Incident* class classifies unwanted behavior in different threat levels. The information in this class is derived from OWL restrictions and rule-based reasoning.

*Device* class includes detailed information about end-devices (e.g., workstations, servers, printers, BYOD) and intermediary devices. End-devices can be mapped to specific entities (e.g., employees, former employees, visitors) and locations (e.g., exact location of a workstation or an access point in a building).

## VI. VALIDATION AND EXPERIMENTAL RESULTS

In this section we validate our system into a real but at the same time controlled environment, present the results, and discuss the limitations and the lessons learned. *PS0* was deployed at the Department of Informatics (IFI) at the University of Oslo in July 2017. Figure 3 presents an architectural floor plan demonstrating the location of available sensors and devices that are part of the experiment.

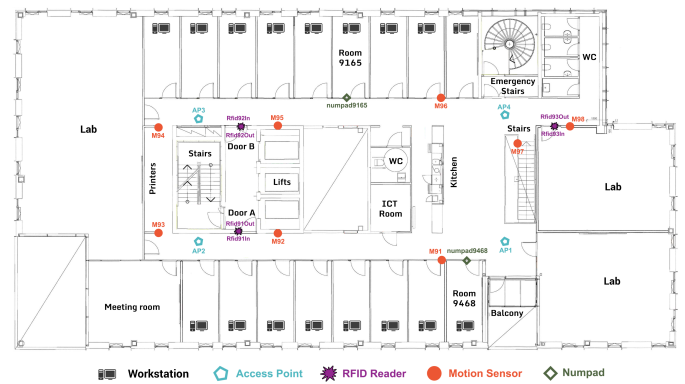

Fig. 3: Architectural floor plan.

To validate the framework and the system we conducted eleven different experiments analyzing ten different use cases (scenarios) of incremental complexity. Each scenario was analyzed by a different student (analyst), currently in the process of acquiring a Master's degree in IT. Furthermore, the eleventh experiment investigated whether acquired domain expertise would decrease the time needed to analyze a future incident considerably. For that reason one analyst went through the process of investigating all ten scenarios one by one. The experiments were conducted for twelve consecutive days; one day for each scenario (experiments one to ten) and two days for the last experiment, eleven. None of the analysts had prior experience with the system. All analysts had an initial one hour training period before the forensic analysis to investigate and get accustomed to the map of the building, as well as all types of sensors and information available for manipulation.

For the sake of space, in this paper we demonstrate the forensic analysis of use case five which is the cutting point in terms of use case complexity and analysis time (Figure 4).

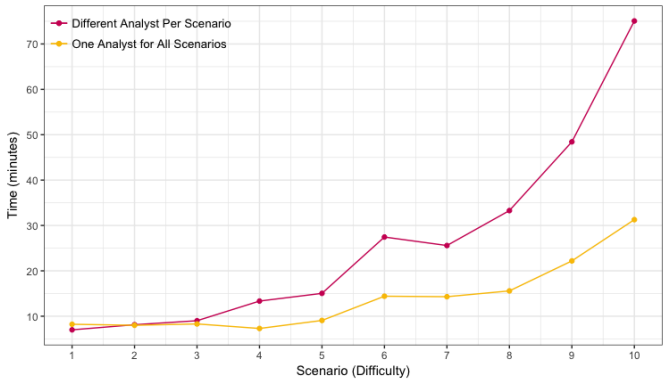

Fig. 4: Use case complexity vs. Analysis time.

### A. Results

All eleven analysts without having any prior experience with the system succeeded to identify the insiders. This is mainly based on the intuitive approach the analysts followed to investigate the incidents. In addition, experiment eleven proved that accustomation to the physical security architecture decreases significantly the time needed to perform the forensic or detection analysis (see Figure 4). The decrease in time is not related to approaches that include prepared queries from acquired domain expertise or the time needed to create the queries, but mainly the time needed to process the available information and come up with an efficient and effective strategy to analyze the information. The queries produce provenance graphs (attack pattern reconstruction) based on the time sequence of the events, thus, providing the appropriate context needed for attack attribution and enriching intelligence.

Limitations of the system in its current state is the complexity of creating the queries and the difficulty of analyzing huge amounts of information (query results) when dealing with big and complex environments. In response to that, prepared queries, graphical user interfaces, or visualization techniques (such as graph databases) can be leveraged to reduce the complexity and support the analysis phase providing us with a full contextual picture of the events and actions generating the alerts.

### B. Use Case Five - Description

The analysts that conducted the analysis did not have any information about the use cases prior to the experiments. Use case five is fully described to improve the readability and understanding of this paper to the reader. *K*, *V*, and *P* are three individuals working at the Department of Informatics of the University of Oslo. *K* is working on the 8th floor and is located in the office *k* with number 8468 ($K \longrightarrow k$). *V* and *P* are working on the 9th floor and they are located in the office *v* and *p* with numbers 9468 ($V \longrightarrow v$) and 9165 ($P \longrightarrow p$), respectively. Access to an office is granted only when an individual successfully authenticates using its private passcode. Both individuals entered the building together from the 2nd level at 22h41 on July 21st using *K's* RFID card. *K* scanned his RFID card once more in the second floor to access the lifts with *V* and again in the 8th floor to access the main space where the offices are located. *K* accessed his office *k* and *V* continued taking the stairs connecting the 8th floor to the 9th floor. When *V* reached the 9th floor he walked towards the targeted office *p* ($P \longrightarrow p$). *V* was able to access the office of *P* after successfully authenticating. It seems that *V* shoulder surfed *P* in the past effectively acquiring the information needed for launching the attack. Eventually, *V* got into his hands a valuable flash drive with available intellectual property (state of the art research). The attacker repeating the same pattern went back to the 8th floor to find *K*. The attacker was aiming to exploit *K's* good intention and use his RFID card to leave the building. Unfortunately, *K* had left his office and the attacker *V* had to use his RFID card to leave the 8th floor and then the building.

### C. Use Case Five - Forensic Analysis

Having knowledge of the building structure, having a clear understanding of the strategic points the sensors were installed, knowing what kind of information the sensors log, and by doing some intelligent analysis we can identify who the attacker is. The conducted forensic analysis does not take into consideration any rule-based abnormality inferred to begin with, but it mainly focuses on manual analysis of the events. We discuss the rule-based reasoning capability approach in the next section. An analyst could take any different approach to investigate the presented use case. The analysis below is one out of many possible. Each incident is unique, thus domain expertise would enable more efficient and effective forensic analysis. This is demonstrated in Figure 4 where each use case difficulty and time needed for completing a forensic analysis is compared among different analyst for each use case (red data points) and one analyst for all use cases (yellow data points).

For the analysis we used SPARQL, a semantic query language able to retrieve and manipulate data stored in Resource Description Framework (RDF). Initially, the only known fact to launch the forensic analysis is that there was a breach in the office 9165. We also know that *P* left his office at 19h24 and accessed it again the following day at 8h32. First we check what time *P's* office was accessed during his absence, as well as any failed attempts.

```
PREFIX ps0:
  ↪ <http://www.semanticweb.org/ontologies/ps0#>
SELECT ?Event ?Date ?Success ?Fail
WHERE {
        ?Event   ps0:hasNumpadSensor ?Numpad .
        ?Numpad  ps0:hasPlace ps0:Room9165 .
        ?Event   ps0:hasTime ?Date .
        OPTIONAL
        {
        {?Event ps0:hasAccessGranted ?Success}
        UNION
        {?Event ps0:hasAccessNotGranted ?Fail}
        }
        FILTER (?Date >
        ↪    "2017-07-21T19:00:00"^^xsd:dateTime
        && ?Date <
        ↪    "2017-07-22T09:00:00"^^xsd:dateTime)
}
ORDER BY ?Date
```

The results of the query show that the office *p* was accessed at 22h45. In addition, we do not see any failed attempts. The analyst decided that the second query should determine if anybody left from the 8th or the 9th floor after the breach. In addition, the query was restricted to a specific time period of seven hours and 15 minutes; from 22h45 to 06h00.

```
SELECT ?Event ?Date ?Location ?Place ?LastName
  ↪ ?FirstName ?Email
WHERE {
        ?Event   ps0:hasTime ?Date .
        ?Event   ps0:hasReadId ?RfidId .
        ?Person  ps0:hasRfidCard ?RfidId .
        ?Person  ps0:hasLastName ?LastName .
        ?Person  ps0:hasFirstName ?FirstName .
        ?Person  ps0:hasEmail ?Email .
        ?Event   ps0:hasRfidSensor ?Sensor .
        ?Sensor  ps0:hasLocation ?Location .
        ?Sensor  ps0:hasPlace ?Place

        {?Event ps0:hasRfidSensor ps0:rfid91In}
        UNION
        {?Event ps0:hasRfidSensor ps0:rfid81In}
        UNION
        {?Event ps0:hasRfidSensor ps0:rfid22In}
        UNION
        {?Event ps0:hasRfidSensor ps0:rfid21In}
        FILTER (?Date >
        ↪    "2017-07-21T22:45:12"^^xsd:dateTime
        && ?Date <
        ↪    "2017-07-22T06:00:00"^^xsd:dateTime)
}
ORDER BY ?Date
```

The results show that an individual named *V* accessed the lifts in the 8th floor at 22h47 and left the building a minute later from the second floor. Since *V* is the only person that left the 8th or the 9th floor after the incident, we request more information regarding the individual's pattern for this particular day. Specifically, we would like to check what time *V* entered the building, if *V* accessed his office, as well as any pattern related to the incident from the logs of access points.

```
SELECT DISTINCT ?Event ?Date ?Location ?AP ?Place
WHERE {
        ?Event   ps0:hasTime ?Date

        {?Event ps0:hasEmail
        ↪    ps0:vasileim\@ifi.uio.no .
        ?Event   ps0:hasAccessed ?MacAddress .
        ?AP      ps0:hasMacAddress ?MacAddress .
        ?AP      ps0:hasLocation ?Location}
        UNION
        {?Event ps0:hasReadId ps0:26964897076 .
        ?Event   ps0:hasRfidSensor ?Sensor .
        ?Sensor  ps0:hasLocation ?Location .
        ?Sensor  ps0:hasPlace ?Place}
        UNION
        {?Event ps0:hasNumpadSensor ps0:numpad9468}
        FILTER (?Date >
        ↪    "2017-07-21T07:00:00"^^xsd:dateTime
        && ?Date <
        ↪    "2017-07-22T06:00:00"^^xsd:dateTime)
}
ORDER BY ?Date
```

Surprisingly in our results there is no information pointing out that *V* accessed the building but interestingly one device accessed the university's network using *V's* credentials. The analyst knowing the security policy of the university assumed that someone gave *V* access intentionally (opened the doors for *V*) or unintentionally (*V* took the opportunity to enter the building when someone else was entering or leaving). Furthermore, the results show an interesting pattern. *V* was in the eighth floor at 22h43, then appeared in the ninth floor (most probably took the stairs), came back to the eighth floor again after the incident, then accessed the lifts by scanning his own RFID card, and finally left the building.

The analyst wanted to identify who gave *V* access and investigate the incident in detail. First, we define a time period of minus-plus 30 seconds from the first identified event related to *V* (the observed remote authentication to an access point close to the entrance of the building in the second floor). Next, we check if in the same time period someone accessed or left the building.

```
SELECT  ?Event ?Date ?Location ?Place ?LastName
  ↪ ?FirstName ?Email
WHERE {
        ?Event   ps0:hasReadId ?RfidId .
        ?Event   ps0:hasTime ?Date .
        ?Event   ps0:hasRfidSensor ?Sensor .
        ?Sensor  ps0:hasLocation ?Location .
        ?Sensor  ps0:hasPlace ?Place .
        ?Person  ps0:hasRfidCard ?RfidId .
        ?Person  ps0:hasLastName ?LastName .
```

```
        ?Person ps0:hasFirstName ?FirstName .
        ?Person ps0:hasEmail ?Email

        {?Event ps0:hasRfidSensor ps0:rfid21In}
        UNION
        {?Event ps0:hasRfidSensor ps0:rfid21Out}
        FILTER (?Date >
        ↪ "2017-07-21T22:41:00"^^xsd:dateTime
        && ?Date <
        ↪ "2017-07-21T22:42:00"^^xsd:dateTime)
}
ORDER BY ?Date
```

The results show that an individual (*K*) actually accessed the building from the second floor during the pre-defined time period. We can confidently assume that *K* gave access to *V*.

### D. Use Case Five - Rule-Based Anomaly Detection

Use case five inferred three indicators that were classified into the ontology as incidents for further investigation and flagged an alert. First, there is information that *P*'s office was accessed but there is no event indicating that *P* entered the building. Second, *V*'s credentials were observed connecting a device to an access point but there is no event indicating that *V* entered the building. Third, *V* was monitored leaving the building but there is no event indicating that *V* actually entered the building in the first place. The above information could be used both proactively and retroactively to mitigate an insider attack or aid an investigation.

## VII. CONCLUSION

In an idealistic environment any type of malicious activity should be prevented or detected and mitigated but this is almost never the case, especially when the attacker is a trusted authority. It is the case that many times malicious activity goes undetected for a long period and incidents are not reported in a timely manner. For example, sabotage would be observed in a short period due to its disruptive nature but espionage could go undetected for a long period. In response to that, we introduced and validated an ontology-based framework (*PS0*) that integrates physical security controls with technical and administrative security controls for improving the security posture of organizations against sophisticated insider physical attacks. Overall the framework supports proactive insider threat detection and investigation, forensic data analysis for attack attribution and thwart deception, reconstructing complex attack patterns for enriching and sharing intelligence, as well as continuous security compliance monitoring.

## ACKNOWLEDGMENT

This research is supported by the Research Council of Norway under the Grant No.: IKTPLUSS 247648 and 248030/O70 for Oslo Analytics and SWAN projects, respectively.